# Identification of Auger mechanisms responsible for low energy electron emission from graphene on copper using Auger-gamma coincidence spectroscopy

R. W. Gladen[1#], V. A. Chirayath[1‡], P. A. Sterne[2], A. J. Fairchild[1], A. R. Koymen[1], and A. H. Weiss[1]

[1]*Department of Physics, The University of Texas at Arlington, Arlington, TX, USA 76114*
[2]*Lawrence Livermore National Laboratory, Livermore, CA, USA 94550*

[‡] corresponding author email: chirayat@uta.edu
[#] corresponding author email: randall.gladen@uta.edu

**Abstract:** We have applied positron annihilation induced Auger-gamma coincidence spectroscopy to identify important mechanisms responsible for the emission of low energy electrons following the sudden creation of holes in bilayer graphene on copper substrate. The novel spectroscopic method measures the energy of the Doppler shifted annihilation gamma photon in coincidence with the Auger electron emitted following the relaxation of the hole created by the annihilation of a surface electron with a surface trapped positron. By extracting and theoretically modelling the annihilation gamma spectra coincident with low energy electrons, we associate majority of the intensity in the low energy (7 eV to 25 eV) region of the Auger spectrum to electron emission following Auger decays of 2s holes in adsorbed oxygen and deep valence holes in graphene. We provide additional support to this conclusion by showing that most of the Auger electrons with energy less than ~ 25 eV are coincident with gamma photons with small Doppler shift (511 keV - 512 keV) indicating that the primary electron whose annihilation resulted in low energy Auger electron emission had small momentum parallel to the gamma emission direction. On the other hand, majority of the Auger electrons measured in coincidence with the photons having maximum Doppler shift (515 keV – 521 keV) were emitted following the decay of core holes (C 1s and O 1s). By selecting annihilation gamma photons coincident only with the O KVV (from surface adsorbed oxygen), C KVV (graphene), and Cu MVV (copper substrate) Auger electrons, we have also experimentally derived the energy spectrum of Doppler shifted gamma photons resulting from the annihilation of positrons with 1s electrons of O, 1s electrons of C and 3p electrons of Cu respectively. Model Doppler broadened gamma spectra produced using ab-initio calculations agree well with the experimentally derived line shapes. This demonstrates the ability of Auger-gamma coincidence method to experimentally resolve the Doppler broadened annihilation gamma spectrum into its veiled electronic level constituents which has, heretofore, relied solely on theoretical analysis.

1. **Introduction**

Auger mechanisms play pivotal role in variety of physical and chemical processes. Auger decay of deep valence or shallow core holes can result in the emission low energy electrons (< 20eV) that are highly efficient in causing biological radiation damage through DNA strand breaks [1, 2]. Multiple localized holes near Fermi level that are left behind at the end of an Auger decay can initiate subsequent processes like ion desorption [3, 4], columbic explosion/decay [5, 6] and can result in the creation of rate determining active reaction sites responsible for catalytic action or molecular fragmentation [7, 8]. Auger cooling of hot holes also influences carrier multiplication [9] and play major role in the "efficiency-droop" of light emitting diodes [10]. However, Auger decay processes that result in low energy electron emission into vacuum are particularly difficult to identify by electron or photon induced electron spectroscopies due to the obscuring presence of secondary electron background. Positron annihilation-induced Auger electron spectroscopy (PAES)—a method that features enhanced surface selectivity with no probe-induced secondary background [11-16]—has recently been used to measure the energy spectra of previously unidentified low energy Auger emission processes. Chirayath et al.,[17] demonstrated that the Auger electron spectrum measured from single layer graphene has a significant contribution from the Auger decay of deep holes (VVV) in the wide valence band of graphene. Similarly, Fairchild et al., [18] provided direct evidence for low energy electron emission following O LVV Auger transitions in $TiO_2$. Both studies relied on (i) the absence of secondary background unrelated to Auger emission processes, (ii) an estimate of the intensity of the inelastic tail of the high energy Auger peaks, and (iii) modelling of the low energy Auger spectrum using the two-particle density of states (equivalent to applying energy-conservation) to identify the initial hole responsible for the Auger emission. However, the information of the energetic hole that leads to Auger emission is carried by the annihilation gamma spectrum in the form of the momentum of the electron with which the positron annihilates. Specifically, the component of the electron-positron pair momentum (dominated by the electron momentum as the positron thermalizes prior to annihilation) parallel to the direction of gamma emission causes a Doppler shift in the energy of the gamma photon leading to a symmetric Doppler broadening of the 511 keV annihilation gamma peak.

 Here, we have utilized the enhanced capabilities of our newly constructed positron beam [19-21] to combine Time-of-Flight (ToF)-PAES with Doppler Broadening Spectroscopy (DBS) to elucidate the Auger relaxation mechanisms that are responsible for the emission of low energy electrons from bilayer graphene overlayers grown on a polycrystalline copper substrate. By selecting and analyzing gamma photons coincident with specific regions of the PAES energy spectrum, we find that majority of the intensity in the low energy region of the Auger spectrum is due to the decay of O 2s and valence band holes, providing support to the findings in [17] and [18]. In addition, we have experimentally extracted the energy spectrum of Doppler shifted gamma photons resulting from the annihilation of positrons with 1s electrons of O and 1s electrons of C. The experimentally derived line shapes compare very well with the theoretical Doppler broadened line shapes obtained using simple atomistic calculations [22]. It should be noted that our group has previously demonstrated the extraction of Doppler line shapes corresponding to Cu 3p and Ag 4p annihilations by measuring gamma photons coincident with a narrow range of $M_{2,3}VV$ and $N_{2,3}VV$ Auger peaks from Cu and Ag [23, 24]. However, in these experiments, the energy of the Auger electrons were measured with a trochoidal energy analyzer, and therefore the Doppler broadening of the annihilation gamma was measured in coincidence with Auger electrons that were constrained to a narrow range of energies — limiting the generalization of the coincidence technique over a wide range of electron energies. In addition, the experimental setup did not allow the reverse coincidence measurements, in which the Auger electron energies are measured in coincidence with specific gamma energies. The recent developments on the University of Texas at Arlington (UTA) positron beam has granted us the ability to measure a broad range of electron flight times in coincidence with a broad range of annihilation gamma energies [19–21], resulting in a two-dimensional (2D herein) histogram that can be used to extract

Auger electron spectra or gamma spectra satisfying specified electron ToF-gamma energy coincidence conditions. We have used this capability to show that the majority of the electrons that are coincident with gamma photons close to the 511 keV gamma peak (i.e. photons with small Doppler shifts) are low energy electrons, whereas the majority of electrons associated with gamma photons in the wing region of the 511 keV gamma line (i.e. photons with large Doppler shifts) are high energy Auger electrons corresponding to core annihilations. Since small Doppler shifts are associated with valence annihilations [25-31], low energy Auger electrons are shown to be associated with the annihilation of low momentum shallow core or deep valence electrons.

Measurement of the energy of the Auger electrons that are emitted following positron annihilation in coincidence with Doppler-shifted annihilation gamma photons of a specific energy giving insights into the primary hole responsible for the Auger decay process is very similar to Auger-Photoelectron Coincidence spectroscopy (APECS) [32]. Conversely, measuring the momentum of the electron with which positron annihilates (through the amount of Doppler shift of the 511 keV gamma photon) in coincidence with the selected Auger electron energies resembles the reverse APECS method where the energy spectrum of the photo electron is measured in coincidence with Auger electrons falling within a selected energy window [33]. However, the ability of PAES to measure low energy (< 25 eV) Auger electron spectrum without the probe or the outgoing valence band photoelectron induced [34] secondary electron background makes annihilation induced Auger-gamma coincidence spectroscopy a distinctive tool for identifying and analyzing important processes that result in low energy electron emission from surfaces. An experimental extraction of the individual orbital annihilation components of the Doppler spectra as demonstrated here can provide a testbed for the ab-initio methods used for calculating material-dependent positron annihilation parameters. Decomposition of the 511 keV annihilation gamma spectrum into its orbital annihilation components provides elemental signature [25, 26] and defect geometry [27-29] at the site of annihilation making it an important component for the interpretation of the positron data [30, 31] and up to this time, was done entirely using ab-initio calculations without experimental support.

In the subsequent sections, we briefly discuss the experimental method used to collect the 2D coincidence data followed by a discussion of the key results that have been derived from it, as well as the future applications of positron annihilation-induced Auger-Gamma coincidence spectroscopy in illuminating physical processes that can occur following an Auger relaxation or during positron-surface interactions [14, 35, 36].

2. **Experimental Methods**

The apparatus used for the experiment consists of a variable energy positron beam, a sample chamber, an electron ToF analyzer with an electron flight path of ~ 3 m, and a high-purity Ge detector (HPGe), which are described in detail elsewhere [21]. The energy of an emitted electron is determined from its time of flight (ToF), which is measured as the time difference (Δt) between the detection of the electron at a micro-channel plate (MCP) detector and the detection of the annihilation gamma by the HPGe detector. The primary advantage of ToF technique is its ability to perform simultaneous and efficient collection of all electrons occupying a wide range of energies. By using the HPGe detector signals as one of the inputs to measure the electron ToF, we were able to obtain both the ToF of the positron-induced electrons as well as the energy of the annihilation gamma photons with just two input signals [19-21].

The sample consisted of bilayer graphene deposited on a substrate of polycrystalline Cu purchased from ACS materials [37]. The presence of graphene on the substrate was verified by established features in the Raman spectra [38]. The sample was biased to -0.5 V, with respect to the ToF analyzer, and a beam of positrons with a mean kinetic energy of ~4 eV was deposited on the sample. It is possible to use such low energy positrons for PAES as the primary hole which undergoes Auger decay is created via annihilation and not by impact. PAES thus, has an additional advantage of obtaining the spectra of Auger electrons from surfaces mostly free from probe-induced secondary electron background. In the present case, the maximum total energy of the deposited positrons is less than 6.5 eV; hence, the maximum kinetic energy of the secondary electrons ($KE_{max}^{e^-}$) that can be ejected from the sample is $KE_{max}^{e^-} = 6.5 - (\varphi_- - \varphi_+) \sim 3.5$ eV, where $\varphi_-$ is the electron work function (~4.5 eV) [39] and $\varphi_+$ is the positron work function (~1.5 eV) [40]. Here, we analyze only electrons with energies greater than 6 eV, which is larger than the maximum energy due to secondary emission processes and, therefore, involves only electrons related to the Auger emission process. Previous investigations of a similar sample have shown the presence of oxygen adsorbed on the graphene overlayer [17], and hence, Auger transitions associated with O along with those from C and Cu are expected in the PAES spectrum

The deposited ~4eV positrons trap in an image potential-induced surface state on the vacuum side of the sample prior to annihilation. The wavefunction of the surface-trapped positrons has limited overlap with electrons beyond the first monolayer and, therefore, signals (Auger electrons and annihilation gamma-rays) emitted following electron-positron annihilation primarily involve surface atoms. This process is responsible for the enhanced surface selectivity of PAES compared to electron- or photon-induced electron spectroscopies. The annihilation gamma rays emitted from the surface were collected by the HPGe detector, and the resulting detector pulses are used for both timing and energy measurements; by unifying these two measurements, the gamma energy of each individual positron-electron annihilation is correlated with the time-of-flight (Δt) of the annihilation-induced Auger electron. This Δt is derived using the analysis of digitally collected signals from the HPGe detector and the MCP detector using internally developed software and a 2D array representing coincident electron ToFs and the annihilation gamma energies is constructed [19, 20].

## 3.    Results and Discussion

The 2D array (i.e. the histogram) consisting of the coincident electron flight times and annihilation gamma energies is represented by the heat map in Figure 2(a). An expanded view of the region around the 511 keV annihilation gamma peak is shown in the heat map. The histogram is comprised of 512 ToF bins × 2048 gamma energy bins, with each cell of the array containing the number of electrons that were both (1) detected within that cell's ToF bin; and (2) were coincident with annihilation gamma photons that were measured to be within that cell's gamma energy bin. If we project all the counts of the 2D histogram onto the ToF axis, we acquire the cumulative positron annihilation-induced Auger electron spectrum given in Figure 2(b); and a projection onto the gamma energy axis yields the Doppler-broadened gamma spectrum of Figure 2(c). The Auger peaks in the PAES spectra that correspond to core annihilations are marked in the cumulative ToF-PAES spectra (Figure 2(b)) with ToFs from 0.9 μs–0.2 μs, corresponding to electrons emitted following Cu $M_{2,3}$VV, O KVV and C KVV Auger transitions. In the heat map of Figure 2 (a) the gamma peak at 511 keV can be seen to be broader in the 0.9 μs–0.2 μs ToF range compared to electron ToFs > 0.9 μs providing an illustrative guide to the relation between the momentum of the electron with which positron annihilated and the energy of the emitted Auger electron. We use this relation between the electron ToF and the Doppler broadening of the gamma spectra to extract Doppler broadened spectra corresponding to positron annihilations with electrons of specific orbital and the method is described below.

The cumulative PAES spectrum comprises of four main regions; these are the spectral peaks corresponding to the Cu $M_{2,3}$VV (0.9 μs–0.65 μs: 58 eV), C KVV (0.65 μs–0.3 μs: 263 eV), O KVV (0.3 μs–0.2 μs: 503 eV) Auger transitions, along with a high intensity in low energy between 6 eV and 25 eV. We can extract the Doppler broadened gamma spectrum due to the annihilation of positrons with O 1s electrons by integrating the counts in the 2D histogram within the range of O KVV Auger electron flight times (highlighted by the orange diagonal stripes in Figure 3(a)). Likewise, the individual spectra of gamma photons emitted following positron annihilation with C 1s and Cu 3p electrons can be extracted by integrating the 2D array over their respective highlighted ToF regions (C: solid magenta; Cu: dashed olive green) in Figure 3(a). These component spectra are given in Figure 3 (b)-(d), respectively. We have extracted only the gamma photons with energies equal to or greater than 511 keV since the lower energy (< 511 keV) region of the gamma spectra contains processes extrinsic to the annihilation processes in the sample (e.g. insufficient charge collection in the detector or positronium formation). We have subtracted the contribution of accidental background from the coincidence gamma spectra by collecting a non-coincident gamma spectrum—in an otherwise identical configuration as the coincidence spectrum—and scaling it by a factor derived from the average background counts in the ToF-PAES spectra in the region beyond the O KVV peak. A comparison of the three extracted gamma line shapes shows that the width of the Doppler broadened gamma peak corresponding to O 1s annihilations is larger than the gamma peak due to C 1s annihilations and that due to Cu 3p annihilations. This is consistent with the trend of increasing Doppler shift with increasing binding energy (~75 eV for Cu to ~540 eV for O) of the electron with which the positron annihilates. However, the Doppler broadened gamma line shapes are much more complex than just its width. In order to verify the efficacy of the new method to obtain orbital level contributions to the total Doppler-broadened spectra, we compared the shape of the experimentally extracted annihilation peaks to the theoretically calculated line shapes (solid lines in the Figure 3(b)-(d)).

The solid lines in Figure 3(b)–(d) representing the Doppler broadened gamma line shapes were generated by an atomistic calculation [22] previously shown to work in cases where the core levels are localized. The annihilation peak shapes of the highly localized core atomic orbitals are primarily influenced by the Coulombic potential of the nucleus and contain very little influence of the specific atomic arrangement on the surface; therefore, the momentum density of the positron-electron pair along one direction of momentum for an electronic state $i$ ($\rho_i(p_L)$) is found by integrating the three-dimensional momentum density, given by Eq. 1, over the other two directions in momentum space:

$$\rho_i(\mathbf{p}) = \pi r_e^2 c \left| \int e^{j\mathbf{p}\cdot\mathbf{r}} \psi^+(\mathbf{r}) \psi_i^-(\mathbf{r}) \sqrt{\gamma(\mathbf{r})} d\mathbf{r} \right|^2 \quad (1),$$

Here, $r_e$ is the classical electron radius, $c$ is the speed of light, $\psi^+$ is the positron wavefunction, $\psi_i^-$ is the electron wave function, and $\gamma$ is a state independent enhancement factor that describes the local enhancement in electronic charge density at the site of positron annihilation. These calculations have been performed using the methods and parametrizations as described in Ref. [22].

    The calculated line-shapes were convolved with a 1.39 keV FWHM Gaussian to represent the instrument response function of the digital HPGe gamma spectrometer at 511 keV [19-21]. The theoretically calculated Doppler broadened gamma line shape corresponding to annihilations with O 1s electrons were scaled by an intensity factor calculated using an error-weighted, least-squares fit [40] to the extracted experimental gamma peak. The calculated annihilation spectrum successfully describes the gamma line shape measured in coincidence with the O KVV Auger electrons (Fig. 3 (b)) giving evidence that all the electrons under the O KVV Auger peak originated following the Auger decay of the O 1s core hole. Our analysis shows that the calculation of the core annihilation (O 1s) line shape using simple atomistic method can completely express the parallel momentum of core level electrons of surface atoms. As seen in the PAES spectra (Fig. 3(a)), inelastically scattered O KVV electrons has little contribution in the region under the C KVV Auger peak. Thus, the annihilation gamma peak extracted in coincidence with electrons under the C KVV Auger peak was successfully expressed using only the theoretically calculated C 1s annihilation line shape (Fig. 3(c)). Here too, we used an error-weighted, one-parameter least-squares fit to obtain the scale factor for the theoretically calculated C 1s annihilation line shape. It should be noted that the region corresponding to electron ToF from 0.9 µs–0.65 µs (38 eV to 72 eV) contains contributions from both the Cu $M_{2,3}$VV Auger peak and the inelastic tail of the C KVV Auger peak. Thus, the gamma line shape coincident with electrons with energies within the 38 eV–72 eV window will have contributions from Cu 3p annihilations as well as C 1s annihilations. Therefore, the experimentally extracted Doppler broadened gamma line shape corresponding to electron ToF from 0.9 µs–0.65 µs (38 eV to 72 eV) was fit using an error-weighted two parameter least-squares method where the coefficients of fit gives the intensity factor for the Cu 3p line shape and the C 1s line shapes. Our fit shows that ~69% of the gamma intensity comprise of annihilations with Cu 3p electrons whereas ~31% of the intensity is from C 1s annihilations. This is consistent with our estimate from the PAES data given in Fig. 3(a). All three fits show the singular ability of the present technique in extracting Doppler-broadened annihilation gamma spectra corresponding to sub-orbital annihilations. It needs to be noted that the energy spectra of C 1s and O 1s annihilation gamma has been extracted experimentally for the first time here.

    Fitting the experimental gamma spectrum, measured in coincidence with electrons falling within a selected energy window, with the calculated Doppler broadened line shapes (as shown in Fig. 3(b)–(d)) demonstrated that it is possible to identify the primary hole whose Auger decay has resulted in the emission of the electrons into the selected energy range. We extend this method to identify the Auger transitions responsible for the electron intensity in the PAES spectrum from 6 eV to 600 eV. Fig. 4 show the high energy region of the annihilation gamma spectra that are coincident with electrons (b) having energy greater than 50 eV, (c) with energy between 27 and 48 eV, (d) with energy between 13 and 21 eV, and (e) with energy between 7 and 10 eV. Fig. 4(a), illustrates the region of the PAES spectra with which gamma spectra in panels 4(b)–4(e) are coincident. One of the first things to note is that the half width at

half maximum(HWHM) of the annihilation gamma peak remains almost similar for the spectra in the first two panels ( ~3 keV and 2.8 keV for Fig. 4(b) and 4(c) respectively) whereas it shows a reduction for annihilation gamma spectra coincident with low energy electrons (~2 keV and 1.7 keV for Fig. 4(d) and 4(e) respectively). The larger HWHM of the 511 keV gamma peak in Fig. 4(b) is due to the fact that all gamma photons contributing to the spectrum are a result of annihilation with O 1s, C 1s and Cu 3p electrons (which resulted in the emission of O KVV, C KVV and Cu $M_{2,3}$VV Auger electrons with energies greater than 50 eV). We checked this by fitting (using least squares method) the gamma spectrum with theoretically calculated Doppler spectra corresponding to O 1s, C 1s and Cu 3p annihilations. As expected, the experimental gamma spectrum in 4(b) could be described completely by these three components with their relative intensities mentioned in the figure legends. The HWHM of the gamma spectra in 4(c) suggest that the annihilation gamma spectrum has similar components as in 4(b) and, therefore, are associated with core annihilations. The gamma spectrum in 4(c) could also be completely described by theoretically calculated O 1s, C 1s and Cu 3p annihilation spectra; the difference from 4(b) being the relative intensity of individual components. Our analysis shows that electrons in the energy window between 27 and 48 eV are mostly inelastically scattered Cu $M_{2,3}$VV Auger electrons with a small contribution from inelastically scattered C KVV and O KVV electrons.

If the electrons in the energy windows between 13 eV–21 eV, and between 7 eV–10 eV of the PAES spectrum were entirely due to inelastically scattered core Auger electrons, the HWHM of the gamma peak will only show a small change corresponding to changes in the relative intensity of the O 1s, C 1s and Cu 3p annihilations. However, the substantial reduction in HWHM of gamma spectra in 4(d) in comparison to those in 4(b) and 4(c) suggest that an appreciable intensity of electrons in the energy window are due to Auger decay of a different and most importantly a shallower (in binding energy) hole. The electron energy range of (13 eV–21 eV) suggest that these electrons cannot be due to VVV Auger electrons as Auger decay of hot valence holes in graphene can result in low energy electrons with maximum energy of ~11 eV [17]. Since we have adsorbed oxygen on the graphene surface, it is possible that the Auger decay of O 2s holes [18] can result in electrons (O LVV Auger electrons) with sufficient energy to be within this energy window. The coincident gamma spectra in 4(d) could be fit by a calculated spectrum that includes the theoretically calculated Doppler broadening spectrum from O 2s annihilations in addition to the gamma spectra from O 1s, C 1s, and Cu 3p annihilations providing proof for the presence of O LVV Auger electrons within this energy window. Considering 21 eV to be the maximum energy of the Auger electrons that can be emitted via an O LVV Auger transition from oxygen on graphene, we estimate that the binding energy of the O 2s level of the oxygen atoms adsorbed on graphene surface is at least ~25 eV and is less than 32 eV. This is consistent with the binding energy of O 2s electrons measured using angle resolved photoelectron spectra collected from oxygen adsorbed on carbon fibers and nanotubes [42].

The gamma spectrum in Fig. 4(e) has a similar HWHM as in 4(d) and, therefore, suggest a similar composition except with different intensities. A fit with similar components (O 2s, O 1s, Cu 3p and C 1s) accurately describes the annihilation gamma line shape coincident with Auger electrons with energies between 7 eV–10 eV. However, Auger emission initiated by annihilation-induced deep holes in the valence band of graphene (VVV) can also result in electrons with energies less than 11 eV, as shown by theoretical simulations in [17]. A fit to the extracted gamma spectrum in 4(e) using C 2s (representing deep holes in valence band of graphene), O 1s, Cu 3p and C 1s also described the line shape effectively. With the signal to background ratio in the wing region of the extracted annihilation gamma spectrum in 4(e), it is difficult to distinguish between the contributions from O 2s and C 2s annihilations. But a further analysis leads us to the conclusion that the majority of the intensity in the low energy region between 7 eV and 10 eV is due to the Auger decay of hot holes in the valence band of graphene and/or shallow O 2s holes. This conclusion is based on (i) the ability of theoretically calculated C 2s and O 2s Doppler broadened line shapes to describe the experimentally extracted gamma spectrum (4(e)) and (ii) due to the fact that, energetically, the Auger decay of hot holes in graphene and O 2s holes can result in Auger electrons with energies of less than 11 eV [17, 18].

The energy of the Doppler-shifted photon is proportional to the momentum of the annihilating electron parallel to the direction of gamma emission; gamma photons having small Doppler shift (energies closer to 511 keV) most likely point to valence electron annihilations whereas photons with large Doppler shift (wing region of annihilation gamma spectra) most probably correspond to core electron annihilations [25-29]. By comparing Auger electron spectra measured in coincidence with photons having small Doppler shift to the electron spectra measured in coincidence with photons having a large Doppler shift, we can thus distinguish between Auger electrons emitted following valence annihilations and those emitted after core annihilations. Fig. 5(a) shows all electrons above 6 eV that were measured in coincident with photons having an energy between 511 to 521 keV. The core Auger peaks of the ToF-PAES spectrum in 5(a) between 0.9 µs–0.2 µs (38 eV to 1250 eV) was fit via least-squares using model ToF spectra representing the O KVV, C KVV and Cu $M_{2,3}$VV, Auger peaks. The model Auger peak included a low energy tail that represented the intensity due to inelastic scattering of the electrons in the Auger peak (estimated using the method suggested by Shirley et al., [43]) as well as electrons emitted via multielectron Auger decay of the primary hole estimated based on previous results from a clean single crystalline Cu sample [15]. The intensity in the ToF-PAES spectra beyond those represented by the core Auger peaks and their low energy tails was fit to an empirical function. We hypothesize that this empirical function represents the energy spectra of electrons ejected because of the Auger decay of hot holes or shallow O 2s holes. Based on the fits it was estimated that 35% of the total intensity shown in Fig. 5(a) consists of electrons emitted as a result of the Auger decay of O 1s, C 1s and Cu 3p holes, whereas the rest of the intensity is due to VVV and O LVV components. The cumulative fit to the ToF-PAES spectrum is shown as a solid line through the experimental data, whereas the core and valence contributions to the Auger spectra are shown as dashed and dash-dot lines, respectively. We subsequently extracted the ToF-PAES spectra (5 (b)) coincident only with photons with energy between 511 keV and 512 keV and the ToF spectra (5 (c)) coincident with photons with energies within the window of 515 keV to 521 keV from the 2D histogram. The variation in the relative intensity of the low energy and high energy region in each ToF-PAES spectrum provides an indication of the nature of the primary hole responsible for the Auger electrons. In Fig. 5(b) the high energy region of the ToF-PAES spectrum (0.9 µs–0.2 µs: 38 eV to 1250 eV) contributes less to the total intensity than in 5(a) and 5(c). A fit using the model line shapes used in Fig. 5(a) to the data shown in Fig. 5(b) shows that core Auger peaks contribute only 27% of the spectrum shown in 5(b) whereas the valence/shallow core annihilations contribute 73%. Similarly, the low energy region of the ToF-PAES spectrum (2.4 µs–1.2 µs) contributes less to the total intensity in 5(c) than in 5(b) or 5(a). A fit using the model line shapes used in Fig. 5(a) to the data shown in Fig. 5(c) shows that core Auger peaks contribute 66% of the spectrum shown in 5(c), whereas the valence/shallow core annihilations contribute only a (highly reduced) 34%. These results give additional support to the hypothesis that the component represented by the empirical function represents Auger electrons emitted following the decay of low momentum valence or shallow core electrons.

Because of the ability of the coincidence technique described above to identify the binding energy of an annihilation-induced core hole involved in a particular event through a measurement of the annihilation gamma energy, many future application of the coincidence techniques described above can be anticipated. One that immediately suggests itself would be to resolve questions regarding physical processes involved in positron annihilation-induced ion desorption from oxide surfaces where the Auger decay mechanisms that result in the emission of neutral or positive ions are ambiguous. Auger-Gamma coincidence can, in principle, lead to an unambiguous identification of the primary positron annihilation-induced hole that initiates the process of ion desorption [34, 35]. Excess energy following positronium formation has been suggested to result in light emission [44] or—if the energetics allow—low energy electron emission through a process similar to Auger neutralization [45], but a definitive proof has not been obtained till now. The identification and the spectroscopic analysis of the light or the electrons emitted via Auger neutralization may result in the development of new tools for investigating electronic properties of materials. The gamma-electron coincidence spectroscopy described here is an ideal tool for extracting spectra of electrons coincident with gamma photons emitted following the decay of Ps atoms.

The potential for future augmentation of the technique also exists: An apparatus capable of measuring the lifetime of the implanted positron (via the use of a pulsed positron beam) in addition to the energies of the coincident electrons and gamma photons can provide definitive chemical signatures of the surface defect states at which positrons annihilate. Furthermore, we have initiated a program in which the experimentally derived gamma line-shapes from standard surfaces are used in conjunction with the theoretically derived line-shapes to train a deep neural network for automatically predicting the chemical composition of complex materials solely from the Doppler-broadened gamma spectra—for example, in inaccessible or internal surfaces. The momentum density distribution of electron orbitals that contribute to the total Doppler spectra extracted using the electron-gamma coincidence method described here will play crucial role during the training of the deep neural network.

## 4.    Conclusion

We utilized a novel experimental method, in which the energy of Doppler-shifted gamma photons is measured in coincidence with the annihilation-induced Auger electrons, to characterize a sample of bilayer graphene on a polycrystalline substrate. By theoretically modelling the annihilation gamma spectra measured in coincidence with the Auger electrons, we show that it is possible to identify the primary holes whose relaxation resulted in the emission of the Auger electrons. Our results show that majority of the intensity in the low energy (high ToF) region of the ToF-PAES spectra is due to both the Auger decay of hot holes in the valence band of graphene (VVV transitions) and the Auger decay of the oxygen 2s holes (LVV transitions). By applying the experimental method "in reverse"—by which the electron ToF spectra corresponding to selected Doppler shift of the annihilation gamma is obtained—provided additional evidence that the intensity seen in the low-energy region of the ToF-PAES spectra is primarily due to the Auger relaxation of holes created by the annihilation of positrons with electrons having small momentum component parallel to the direction of gamma emission. Thus, Auger-Gamma coincidence technique is an ideal tool to elucidate Auger decay of energetic holes that are difficult to be identified using other spectroscopic techniques.

In addition, using the Auger-Gamma coincidence method, we were able to extract the spectra of gamma photons emitted following the annihilation of positrons with (i) 1s electrons of C atoms of bilayer graphene; (ii) 1s electrons of O atoms adsorbed on bilayer graphene; and (iii) 3p electrons of the Cu substrate. The experimentally extracted gamma spectra corresponding to sub-orbital annihilations could be completely described with theoretically calculated line-shapes and, therefore, have helped to benchmark the ab-initio calculations. Our analysis shows that simple atomistic calculation which does not consider the complex surface structure can successfully produce the Doppler Broadened gamma spectrum originating from surface annihilations.

**Acknowledgements:** We gratefully acknowledge the support by Welch Foundation (grant No. Y-1968-20180324) for the support during this work and for the development of the coincidence methods described in the manuscript. The positron beam system at UT Arlington was developed using the NSF major research instrumentation grant DMR-1338130 and NSF grant DMR-1508719. This work was prepared in part by LLNL under Contract DE-AC52-07NA27344.

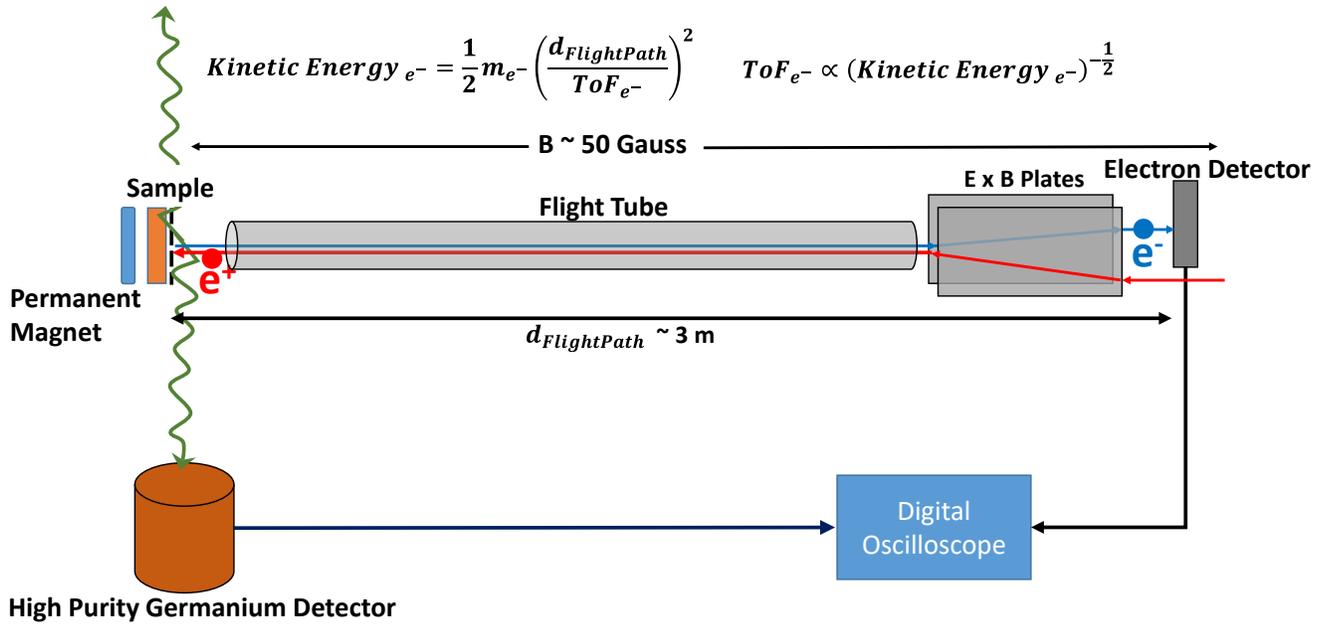

**Figure 1**. **Auger-Gamma Coincidence Spectrometer:** Schematic of the 3 m time-of-flight (ToF) spectrometer in the advanced positron beam at UTA capable of measuring the energy of the Doppler shifted gamma photons in coincidence with the ToF of the electrons emitted following positron implantation. The kinetic energy of the electron can be obtained from its ToF as given in the equations. In the present experiment, positrons with mean energy of ~ 4 eV were magnetically (~ 50 Gauss) transported to a sample that was biased to -0.5 V with respect to the flight tube. The electrons emitted following positron annihilation travel back through the flight tube and drifts upwards towards the electron detector due to the transverse electric ($\bar{E} \times \bar{B}$ plates) and magnetic fields near the electron detector. Annihilation gamma photons were detected using a high purity germanium detector (HpGe) and the signals produced triggered the digital oscilloscope. The digital scope also recorded the electron detector signal arriving within 5.25 μs of the trigger. The coincident signals were used to extract the ToF of the electrons whereas the amplitude of the HpGe signals provided the energy of the Doppler shifted gamma photons resulting in a 2D array of correlated electron ToFs and annihilation gamma energies.

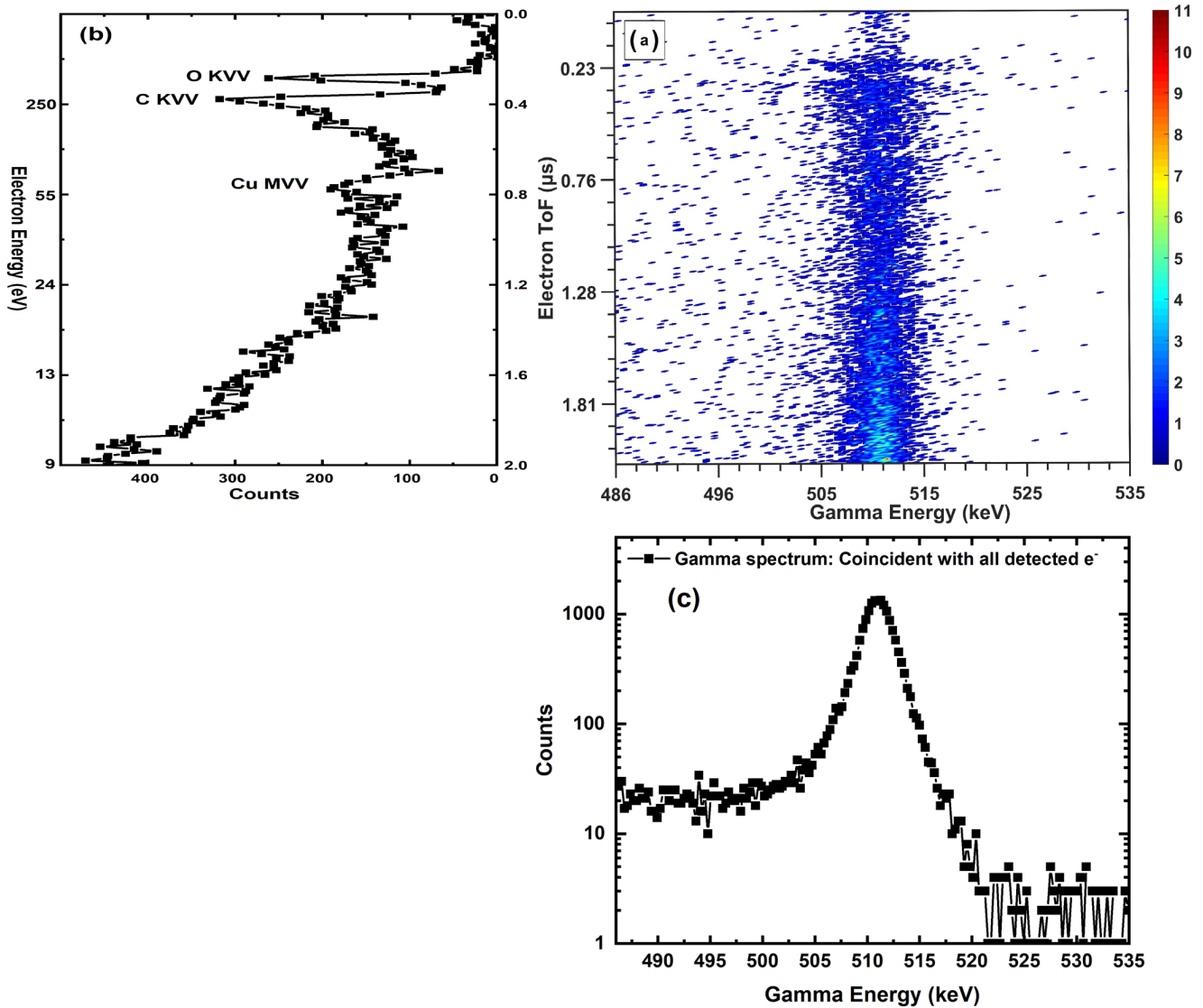

**Figure 2. Two-dimensional representation of the Auger-Gamma coincidence data:** (a) Heat map representation of the two-dimensional array containing correlated electron ToF (vertical axis) and energies of Doppler shifted gamma photons (horizontal axis). An expanded view of the region around the 511 keV annihilation gamma peak is shown. The histogram is comprised of 512 ToF bins × 2048 gamma energy bins. (b) The cumulative Auger electron ToF spectrum obtained by projecting the counts in the 2D array onto the ToF axis. The Auger peaks corresponding to electrons emitted following Cu $M_{2,3}$VV, O KVV and C KVV Auger transitions are marked. The energy corresponding to the ToF is shown in the axis opposite to the ToF axis. (c) The cumulative Doppler-broadened gamma spectrum obtained by the projection of 2D array onto the gamma energy axis by summing along all electron ToFs.

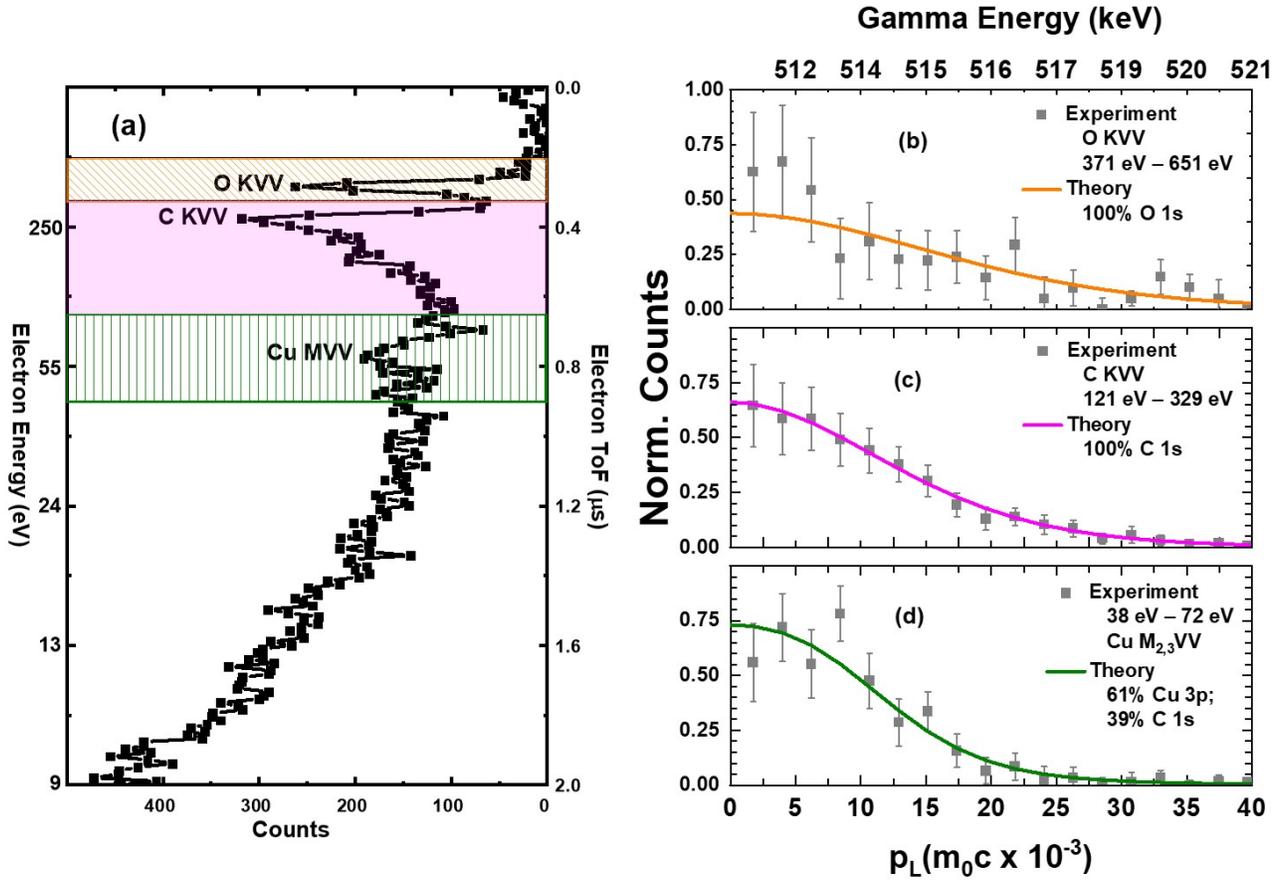

**Figure 3. Gamma Spectra coincident with core Auger electrons:** (a) The ToF-PAES spectra highlighting the range of electron ToFs (or equivalently electron energies as indicated on the left axis) over which the counts in the 2D histogram was integrated to obtain the coincident gamma spectra. Orange diagonal stripes indicate electrons emitted following decay of annihilation induced O 1s holes (O KVV Auger peak). Solid magenta indicate ToF region occupied primarily by electrons emitted following the decay of annihilation induced C 1s holes (C KVV) and olive green stripes represents ToF region where the intensity is mostly due to the decay of Cu 3p holes (Cu $M_{2,3}$VV Auger peak). (b) The experimental spectrum of Doppler shifted gamma photons measured in coincidence with the O KVV Auger peak. The spectrum is fit (solid orange line) with the theoretically calculated one-dimensional electron-positron momentum density distribution (shown as solid line) obtained when positron annihilates with O 1s electrons. The fit is performed using an error weighted least squares method. (c) The Doppler broadened annihilation gamma spectrum measured in coincidence with electrons under the C KVV Auger peak and fit (solid magenta line) with the theoretically calculated electron-positron momentum density distribution when positron annihilates with C 1s electrons. (d) The spectrum of Doppler shifted annihilation gamma photons measured in coincidence with the electrons under the Cu $M_{2,3}$VV Auger peak. The annihilation gamma spectrum is fit with the theoretically calculated one-dimensional momentum density distribution of the positron-Cu 3p electron pair and the one-dimensional momentum density distribution of positron-C 1s electron pair. Our fit shows that the 69% of the Doppler broadened spectrum consists of photons emitted due to the annihilation of positrons with Cu 3p electrons and 31% of the gamma spectrum consists of photons due to positron-C 1s electron annihilation.

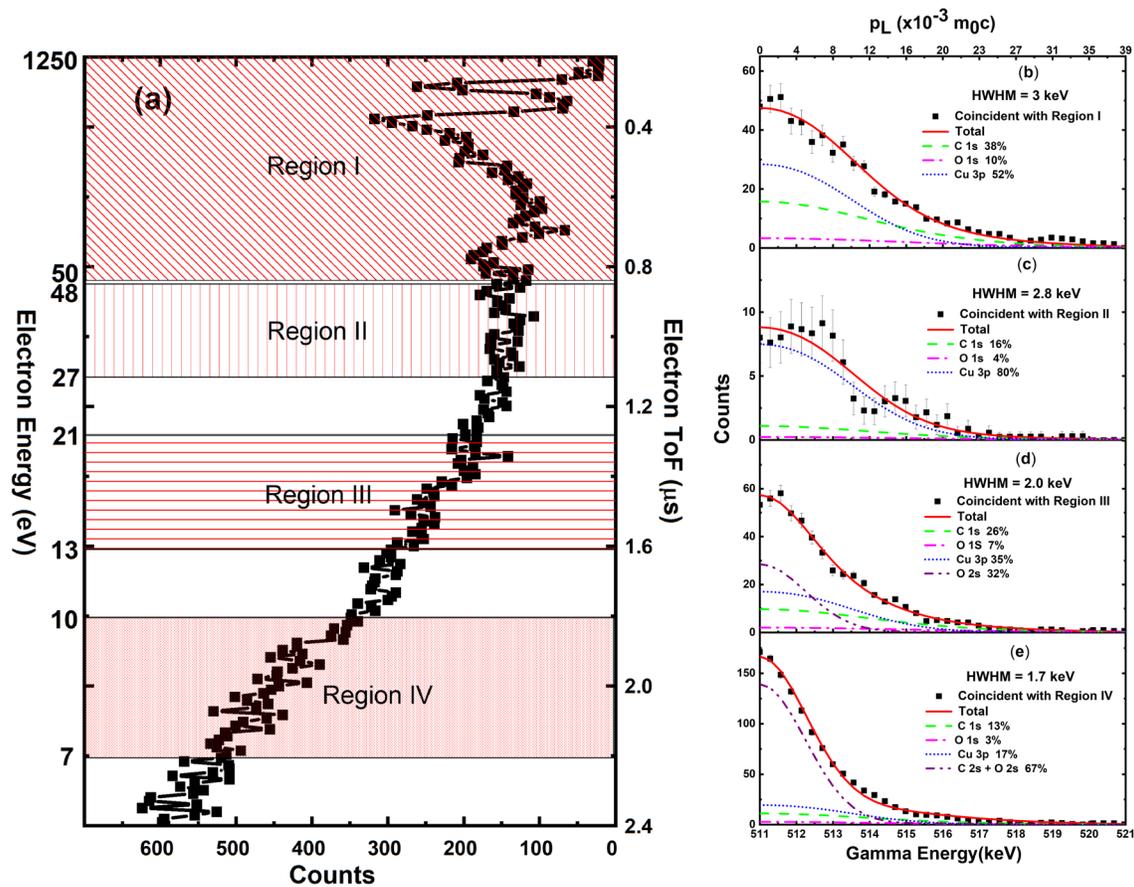

**Figure 4. Gamma Spectra coincident with Auger electrons within selected electron energy windows:** (a) The ToF-PAES spectra highlighting the range of electron ToFs (or equivalently electron energies as indicated on the left axis) over which the counts in the 2D histogram was integrated to obtain the coincident gamma spectra. Diagonal stripes indicate Region I, vertical lines indicate Region II, the horizontal lines represent Region III and solid shades points to Region IV of the ToF-PAES spectrum. (b) The experimental spectrum of Doppler shifted gamma photons measured in coincidence with the core Auger electrons in Region I. The spectrum is fit with the theoretically calculated Doppler broadened annihilation gamma spectrum of O 1s (dash-dot), C1s (dash) and Cu 3p (short-dot) annihilations. The width of the half of the gamma peak shown here at half the maximum value (HWHM) is also given. (c) The Doppler broadened annihilation gamma spectrum measured in coincidence with electrons under the Region II. The spectrum has been fit with the theoretically calculated Doppler broadened annihilation gamma spectrum of O 1s (dash-dot), C1s (dash) and Cu 3p (short-dot) annihilations. (d) The spectrum of Doppler shifted annihilation gamma photons measured in coincidence with the electrons represented by Region III. The spectrum can only be fit by including the O 2s (dash-dot-dot) annihilation gamma spectrum in addition to the core annihilation components (O 1s, C1s and Cu 3p). (e) The spectrum of Doppler shifted annihilation gamma photons measured in coincidence with the electrons in Region IV. The gamma spectrum could be fit with four components (O 2s/C 2s, O 1s, C 1s and Cu 3p). The fit with C 2s and O 2s gave very similar fits and with the present signal to background ratio, it was not possible to distinguish between C 2s and O 2s annihilations.

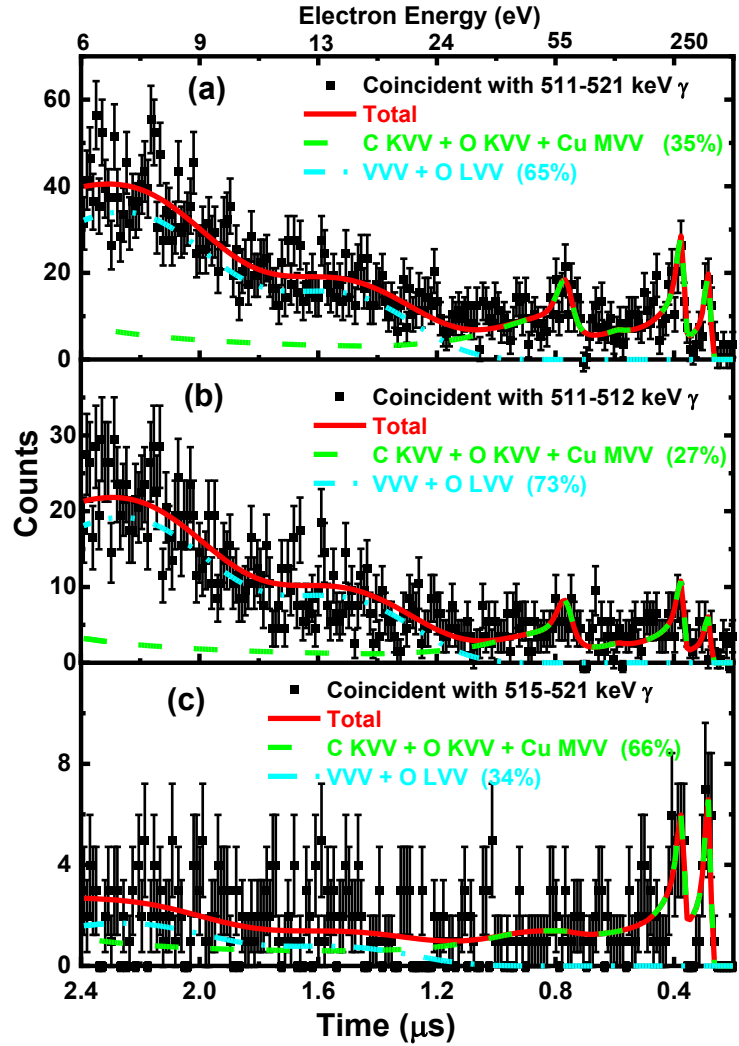

**Figure 5. ToF-PAES Spectra coincident with Doppler shifted gamma photons within selected gamma energy windows:** (a) The ToF-PAES spectra representing all electrons measured in coincidence with annihilation gamma photons with energies between 511 keV – 521 keV. The spectrum is derived from the 2D coincidence spectrum by integrating along the high-energy region of the 511 keV peak. The spectrum is fit using model Auger line shapes (red solid lines) and the relative intensity of the Auger electrons emitted following the decay of the core holes (C KVV, O KVV and Cu $M_{2,3}$VV: green dashed line) and those emitted following the decay of deep valence or shallow core (VVV/LVV: cyan dashed dot dot) obtained following the least squares fit is given. (b) ToF-PAES spectrum obtained by integrating the counts in the 2D coincidence spectrum within 1keV from the 511 keV peak. The relative contribution of the high energy core Auger peaks (green dashed line) and low energy Auger electrons (cyan dashed dot dot) to the total spectra (red solid lines) has been obtained after the least squares fit. (c) The ToF-PAES spectra derived from the 2D coincidence spectrum by integrating along the high-momentum region of the 511 keV peak (516 keV–524 keV). The spectra show large increase in relative intensity of the core Auger peaks (green dashed lines) and a decrease in the relative intensity of the low energy electrons (cyan dashed dot dot) obtained through least squares fit. This provide evidence (complementary to what was shown in [7] and [8]) that these low energy electrons are Auger emissions following the annihilation of positrons with deep valence (VVV) or shallow core (LVV) electrons.